# Composition-independent temperature induced by swift heavy ions in insulators and its relation with track formation

G. Szenes

Department of Materials Physics, Eötvös University,

P.O. Box 32, H-1518 Budapest, Hungary

email: szenes.gyorgy@ttk.elte.hu

**ABSTRACT**

A systematic study of the $R_e$-$T_m$ relation ($R_e$ - track radius, $T_m$ - melting temperature) in insulators reveals new features of track formation. A quantitative relationship is demonstrated between apparently independent $R_e$, $T_m$ data pairs measured in various solids when $<s_e>$=constant ($<s_e>$ - atomic stopping power). This effect is the consequence of a composition-independent, identical maximum temperature distribution $\Delta T(r,0)$ which is induced in various insulators in a narrow cylindrical volume close to the tracks. It has a Gaussian shape, with a width w=4.5 nm, which is independent of materials parameters. It can be measured experimentally by determining the $R_e$, $T_m$ values in various insulators when $<s_e>$=constant. The consequences of the experimental $\Delta T(r,0)$ distribution are discussed. The parameters of the underlying mechanism do not depend on the composition of the solid. The different response of semiconductors is also briefly discussed.

# I. INTRODUCTION

The interaction of energetic ions and solids has great theoretical and practical significance, and has been the subject of intense research for several decades. As a result of ion irradiation, various defects are formed in the solids. It is a minimum requirement for the formulation of any theory that all characteristic features of the defect structure in question must be revealed in appropriate experiments. It is evident that incomplete information in this respect may lead to incorrect theoretical results. In this paper this kind of situation is presented.

It is well-known that the energy deposition depends on the composition of the target. However, experiments show that the induced temperature itself may be not sensitive to the composition of insulators when $<s_e>$=constant ($<s_e>$ - atomic stopping power). Thus the progress of the formation of the induced temperature can be separated from the complete process in this respect. In this paper, the consequences of such composition-independent temperature induced by swift heavy ions are discussed. This kind of temperature was derived unambiguously in the analysis of track formation in insulators [1].

Effects that do not depend on the composition are rather rare in physics even if one limits their study to a group of solids. Except the universal law of gravitation, other analogous effects are not widely known. Additionally, this physical effect is induced by projectiles having velocities up to 0.1c, which generate extremely strong electromagnetic field, leading to formation of intense deformation waves, evolution of a high number of excited energetic electrons and highly ionized lattice atoms. Therefore, the formation of an identical temperature distribution is truly unexpected in this case.

Below, experimental results on tracks induced by high energy ions in crystalline insulators are reviewed. It is shown that presently crucially important information is missing from the experimental basis of various models. In fact, the quantitative relationship found between track sizes measured in different insulators [2] for $<s_e>$=constant has been ignored. This relationship is of major importance for the mechanism of track formation.

Due to this omission, it has not been realized that the mechanisms applied in various theories up until now are not suitable for explaining correctly the formation of tracks even if the popular electron-phonon mechanism of energy transfer is considered. Additionally, various materials parameters (MP) are applied which are actually ineffective for track formation as this is clearly demonstrated in this paper. As a result, important processes might remain hidden in the analyses of irradiation experiments. It is the aim of this study to discuss this problem in details.

Several theoretical models have been proposed for track formation (for a review see [3]) but the most widely used models are the thermal spike models. In these models, the formation of a cylindrical time-dependent high-temperature region is assumed in the irradiated crystal along the trajectory of the energetic ion, where an amorphous phase may be formed from the melt as a result of the high cooling rate. The maximum diameter of the amorphous track $2R_e$ is taken to be equal to the width of the induced maximum temperature increase $\Delta T(r,0)$ (t=0) at the local temperature $T=T_m$ ($T_m$ – melting temperature). The track size is controlled by the electronic stopping power $S_e$ and $R_e>0$ for $S_e>S_{et}$, where $S_{et}$ is the threshold electronic stopping power.

The Analytical Thermal Spike Model (ATSM) is a phenomenological model [4]. A Gaussian distribution is assumed for $\Delta T(r,0)$ in the lattice, whose parameters are determined from the analysis of track experiments. Quantitative relations are found between the responses of various insulators to ion irradiation [2]. This paper is devoted to the analysis of experiments of this type of effects.

## II. THE EXPERIMENTS

Various processes excited by the energy deposition may affect track formation. The most useful experiments for theoretical considerations are those, where the initial track radii $R_{in}$ (t=0) are accessible for measurements because of their direct relationship with $\Delta T(r,0)$. It is reasonable to study the relationship between the irradiation parameters, the properties of the target and the parameters of tracks when systematic $R_{in}$ data are known. However, initial radii are often not accessible, as they may be modified within a very short time by various superimposed processes of phase transformation and recrystallization [5]. Presently, suitable experimental $R_{in}$ data are available only for about 20 insulators and this number has been increased only by one during the last decade.

Unexpectedly, the experiments on track formation show that there is a simple quantitative relationship between $R_{in}$ values when $<s_e>=S_e/N=$constant (N – number density of atoms) even if they are induced in different insulators by different ions of different energies. Such relationship is shown in Fig. 1, where tracks were measured both at low (E<2 MeV/nucleon LO) and high (E>8 MeV/nucleon HI) ion velocities. For symmetry reason, the X-axis extends to negative values in the figure.

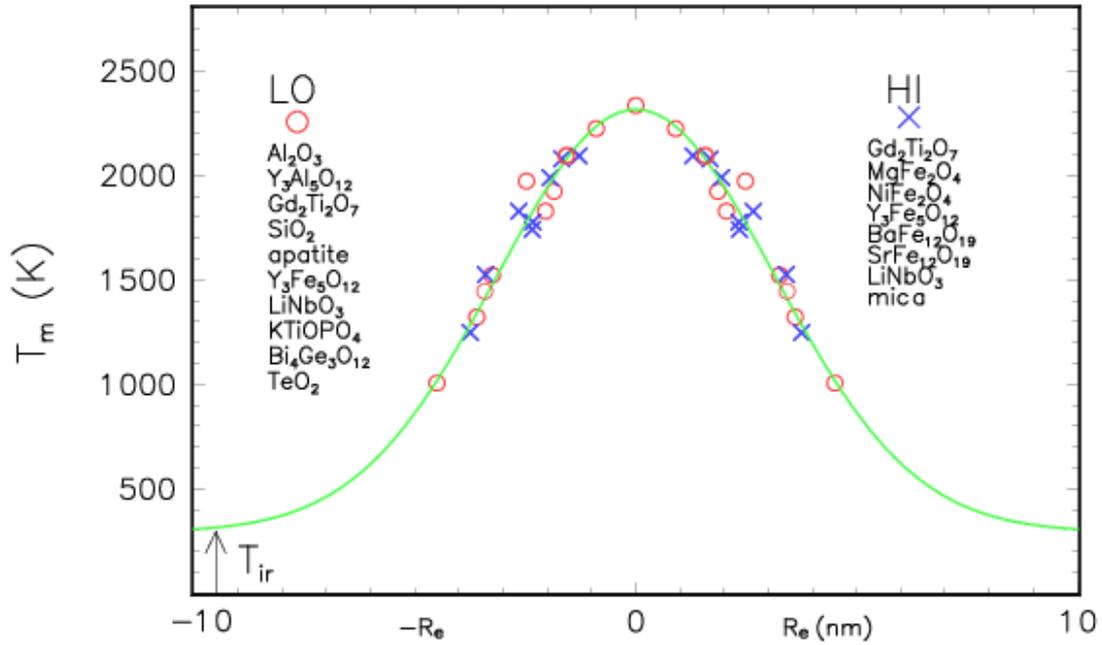

Fig. 1. Variation of track radii $R_e$ with the melting temperature $T_m$ in insulators irradiated by low (o) (E<2 MeV/nucleon) and high velocity (x) (E>8 MeV/nucleon) ions at $<s_e>/3k=3.42 \times 10^5$ nm$^2$ K and $7.5 \times 10^5$ nm$^2$ K, respectively, ($<s_e>$ - atomic stopping power, k – Boltzmann constant, $T_{ir}$ - irradiation temperature, 295 K). In the figure, the solids are listed in the order of decreasing $T_m$. The enveloping curve for $T>T_{ir}$ is $\Theta(r)=T_p\exp\{-r^2/w^2\}$ with w=(4.5 ± 0.3) nm and $T_p$=2020 ± 60 K (for track data see [1,6]).

Though, the smooth curve through apparently independent track data in Fig. 1 is rather astonishing, however, it is not reasonable to consider it as the result of occasional coincidences. Such relationship is rather the consequence of fundamental underlying processes.

The quantitative relationship between track radii in Fig. 1 is very informative and points to the weakness of the currently applied popular theories [7]. The enveloping curve is $\Theta(r)=T_p\exp\{-r^2/w^2\}$ for $T>T_{ir}$, satisfying $r=R_e$ whenever the local temperature is $\Theta(R_e)+T_{ir}=T_m$ ($T_{ir}$ – irradiation temperature, 295 K). In the figure, the two $<s_e>$ values were chosen so that the peak temperatures $T_{pL}$ and $T_{pH}$ be equal for LO and HI irradiations. The systematic relationship between track radii induced in independent experiments was unexpected, but it was predicted by ATSM, together with the correct numerical values of the parameters $w$ and $T_p$ [2]. Previously, predictions of the thermal spike models were expected to deviate from experimental values by a factor of 2 to 4 due to the „enormous simplifications" of these models [8]. This is not that case.

The most important feature of the plot in Fig. 1 is that all conclusions follow directly from experimental data thus their reliability does not depend on the validity of any theoretical assumption. This is important, because a number of various MPs are considered affecting the track sizes, based on theoretical considerations, whose validity has not been checked under non-equilibrium irradiation conditions.

The most important conclusions derived from the plot in Fig.1 are the following.

i) According to Fig. 1, track radii depend only on a single MP - $T_m$. This is the most important evidence for the validity of the thermal spike mechanism of track formation. Other MPs may have only minor effect, otherwise there would be large deviations from the smooth curve in Fig. 1. It has been demonstrated not only for $<s_e>$=constant but in a broad range of $S_e$ (6-54 keV/nm), as well, that the $R_e^2$-$S_e$ track evolution depends only on $T_m$ [6].

ii) The enveloping curve in Fig. 1 is a universal curve [9] in the sense that $T_p$ and $w$ do not depend on MPs.

iii) The same w=4.5 nm appears in the $R_e^2$-$S_e$ track evolution curves as $w^2=dR_e^2/d\ln S_e$ [4] and as the Gaussian width of the $\Theta(r)=T_p\exp\{-r^2/w^2\}$ function in Fig. 1, and its value does not change with the ion velocity. As shown in Ref.[1], there is a simple relationship of the w parameter with basic length units: the Bohr radius $a_0$, the Compton wavelength $\lambda_e$ and the fine structure constant $\alpha$ [1]. Actually,

$$a_0 = \frac{\lambda_e}{2\pi\alpha} \approx \frac{\pi\alpha}{2} w \qquad (1)$$

providing w=4.62 nm in excellent agreement with the experiments. This may be considered as an indication that the existence of the $\Theta(r)$ relation is the consequence of basic interactions which have not been taken into account previously.

iv) The relationship between $R_e$ track radii implies a relationship between the temperatures, induced in different insulators when tracks are formed. If the width of the induced temperature distribution is $R_w$ in a solid at $T_w$ and further experiments in other solids provide other $R_w - T_w$ data pairs then it is reasonable to expect a random $R_w - T_w$ relationship. In Fig. 1 the result of such experiment is shown for $<s_e>$=constant where $R_e$ and $T_m$ are the variables, and instead of a random distribution the $\Theta(r)$ function describes the relationship. This is possible only in that case if the induced temperature distributions $\Delta T(r,0)$ are identical in every solid. The enveloping curve in Fig. 1 provides the common $\Delta T(r,0)$ curve, whose width is equal to the actual track diameter $2R_e$ at each melting temperature. It can be identified as the common induced maximum temperature increase $\Delta T(r,0)=\Theta(r)=T_p\exp\{-r^2/w^2\}$ for the given value of $<s_e>$, that does not depend on MPs. In other words: a composition independent maximum temperature distribution is formed in these experiments [1]. It can be obtained experimentally by using track data of a sufficient number of insulators providing its widths at various $T_m$ temperatures. This experimental curve is in excellent agreement with that assumed in the first publication of the ATSM [4].

This last conclusion becomes evident when one takes into account that the temperature varies in the molten volume along its diameter in the range $T_m<T<T_p$. If one finds that $R_e$ does not depend on MPs, then this means that the width of $\Delta T(r,0)$ does not depend on MPs in the range $T_m<T<T_p$. As this is valid for different solids, $T_m$ is to be considered a variable here. This leads to the conclusion that $\Delta T(r,0)$ must be independent of MPs in a broad temperature range. Thus they are identical in various insulators for $<s_e>$=constant. This explains the accurate quantitative relationship between ion-induced track sizes in different insulators.

v) When $T_m$ values are known, the $R_e^2$-$S_e$ track evolution curves of any of the materials in Fig. 1 can be calculated within experimental error. It is sufficient for this if the experimental $R_e$ data of one of the insulators are measured at different $S_e$ values [9]. Other MPs beside $T_m$ are unnecessary for the calculation.

vi) In Fig. 1, two different values of $<s_e>$ lead to identical $\Delta T(r,0)=\Theta(r)$ distributions with identical width and thermal energy. This is an indication that only a fraction of $<s_e>$ is transferred to thermal energy related to tracks and these fractions are different in the LO and HI ranges. This is the origin of the velocity effect (VE) [10], leading to higher thermal energy in the given insulator and the formation of larger tracks in the LO range for the same value of $S_e$. An expression for the maximum peak temperature $T_p$ can be obtained by using the equation of the balance of energy. Accordingly, a fraction of the deposited energy $fS_e$ (f – efficiency) is equal to the thermal energy leading to

$$T_p = \frac{f<s_e>}{3\pi k w^2}, \qquad (2)$$

where the $\rho c=3Nk$ approximation is applied for the c, the specific heat ($\rho$ – density, N - number density of atoms, and k - Boltzmann constant). The final form is

$$\theta(r) = \frac{f\langle s_e \rangle}{3\pi k w^2} e^{-\left\{\frac{r^2}{w^2}\right\}}. \tag{3}$$

The value of the efficiency f can be estimated from the comparison with the experimental data, and f≈0.4 and f≈0.17 are obtained for LO and HI irradiations, respectively [11].

vii) It is important to note that the condition $\langle s_e \rangle$=constant is required in Fig. 1 for the proper comparison of track radii. This is the direct consequence of the thermal origin of track formation as $T_p$ is proportional to $\langle s_e \rangle$ and $\langle s_e \rangle$=constant guarantees the identical conditions for track formation.

The plot in Fig. 1 is coherent with another type of plot when more data is available for checking Eq. (3). By applying $\Theta = T_m - T_{ir}$ and $r = R_e$ in Eq. (3), it is predicted that in an $R_e^2$-ln ($\langle s_e \rangle / 3\pi k w^2 (T_m - T_i)$) plot the same two parallel lines with slope $w^2$ are obtained for track data of various insulators, and the values of f the efficiencies determine the crossing points on the X-axis at $R_e$=0. This kind of plot was used in several publications as an evidence of the uniform behavior of insulators at LO and HI irradiations in a broad range of $\langle s_e \rangle$ [6]. It is emphasized that no individual fitting parameters are used for different solids in these figures. Nevertheless, the track data follow common curves for various insulators within experimental error. The good agreement with the experiments of the above plots unambiguously confirms the validity of Eq. (3) and the close quantitative relationship between track formation in various solids.

Recently, electronic sputtering data was analyzed in ten solids (insulators and semiconductors) by applying Eq. (3) for the induced temperature [12]. Thermal activation mechanism was assumed [13] and excellent agreement was found with the measured sputtering yield data in a broad range of $S_e$ for all solids. In all previous analyses of sputtering experiments agreement with the calculations could be obtained only in a rather narrow range of $S_e$ in spite of applying individual fitting parameters. The successful application of Eq.(3) is a unique result and it proves convincingly the significance of Eq. (3) for the understanding of various irradiation effects.

The relationship in Fig. 1 and the considerations thereafter confirm unambiguously that in the process of ion-induced track formation, the induced maximum temperature distribution does not depend on MPs, and it is equal to the universal distribution $\Delta T(r,0)=\Theta(r)$ given by Eq. (3). In other words, a composition independent temperature distribution is formed. It is obvious that a temperature distribution which is given by a universal function free of MPs cannot be the result of any of the presently applied theories of the irradiation effects except ATSM.

We emphasize that the calculation of the ion-induced temperature field is not an abstract theoretical question, but a markedly practical problem. Besides track formation and electronic sputtering, the reliable knowledge of the induced temperature is crucial for following the processes of formation of surface hillocks, ion-beam mixing, electronic desorption, stability of radiation defects, as well as for medical applications. Unfortunately, it is impossible to check the estimated temperatures directly, and this leads to uncertainties.

## A. SOME PRACTICAL ASPECTS

Note that a lot of experimental information has been simply ignored in the publications in this domain when they were in conflict with the currently fashionable theories. This is the origin of much confusion concerning the effect of various MPs on the irradiation induced temperatures. Below, some cases are reviewed when ignoring the experimental information shown in Fig. 1 and partially known already in 2007 [2], led to evidently incorrect conclusions, affecting the estimates of the induced temperature. The various applications of the inelastic thermal spike (i-TS) model [3] are the most suitable for the demonstration. Following conclusion i) above, models which do not apply the thermal spike concept are not considered here.

Below some assumptions of i-TS model are compared with the consequences of the plot in Fig. 1. An important element of the phase transition of crystalline materials is the heat of fusion L. In insulators, $L/Q_m$ varies in a broad range, where $Q_m$ is the heat necessary to raise the temperature from room temperature to $T_m$; e.g. $L/Q_m$=0.19 ($SrFe_{12}O_{19}$), 0.44 ($LiNbO_3$), 0.6 ($TeO_2$), 0.66 ($Gd_2Ti_2O_7$). It is obvious from Fig. 1 that L may have only a minor contribution if any, in the process of track formation. Otherwise, one would see a considerable scatter in the figure, since L is an MP, whose magnitude is not proportional to $T_m$. Similarly, this large variation of $L/Q_m$ ought to lead to a considerable scatter and shift of threshold values $<s_{et}>$ in the LO and HI ranges, and this is not observed, either [6]. Thus a systematic study does not confirm any impact of L on track formation [11] while this is a basic element of the i-TS theory [3]. Thus, in this case the contradiction does not occur between two theoretical assumptions but between a firm experimental fact and a theoretical assumption.

Ion irradiation induces a very fast crystalline-amorphous phase transformation. It seems to be reasonable assuming a superheating process. In this process, the phase transition occurs at $T_{ps}>T_m$, where $T_{ps}$ is the superheating temperature, while an additional thermal energy covers the heat of fusion L. According to theoretical estimates, the degree of superheating depends on MPs and varies between $(0.1-0.5)T_m$ regardless of heating rates [14], and thus no proportionality is expected between $T_m$ and $T_{ps}$. The results in Fig. 1 clearly demonstrate that track formation is characterized by $T_m$ and not by $T_{ps}$, and that the contribution of L is completely absent. Consequently, superheating is not induced by ion irradiation.

A typical feature of track formation is VE [10] which means that low velocity ions induce larger tracks than high velocity ions with the same $S_e$. A widely accepted explanation is that compared to low energies, HI ions form broader spatial distribution of excited electrons leading to lower energy density and a broader induced temperature distribution. The result is the formation of smaller tracks at HI velocities at $S_e$=constant. However, it is evident from Fig.

1 that it is not the width of $\Delta T(r,0)$ but its thermal energy varies with the ion velocity. Thus the above explanation is immediately refuted by experimental data (for a more detailed analysis see [11]).

It has been also proposed that track formation is sensitive to the band gap energy $E_g$ [15], the main evidence being the behavior of tracks in $BaFe_{12}O_{19}$ [16]. Again the plot in Fig. 1 shows that small or high values of $E_g$ are indifferent to $R_e$. Actually, $E_g=1$ eV for $BaFe_{12}O_{19}$ and $E_g=12$ eV for $SiO_2$ in Fig. 1, nevertheless, track radii of these two solids fit equally well to the curve $\Theta(r)$ (for a more detailed analysis see [11]).

In various theories, it is always assumed that the total $S_e$ is effective for track formation. However, in the analysis of the plot in Fig. 1, it is noted under vi) that only the $fS_e$ fraction (f<1) is responsible for $\Delta T(r,0)$ and the formation of ion-induced amorphous tracks. This is quite necessary, as the $(1-f)S_e$ fraction covers the energy for inducing other types of radiation defects and for various other processes. It is the origin of serious mistakes when f<1 is ignored. It is emphasized again that the above considerations are not the results of speculations but they are conclusions, which are confirmed by direct experimental evidences.

Though the relationship shown in Fig. 1 and the application of the i-TS model lead to diametrically opposite results, it is remarkable that track data of $Y_3Fe_5O_{12}$, the emblematic solid of i-TS model, nicely fits to the enveloping curve in Fig. 1. According to v) the $R_e^2 - S_e$ track evolution curve of $Y_3Fe_5O_{12}$ can be simply calculated from the data of any other solid in Fig. 1 when only $T_m$ values are known from MPs. This was demonstrated in Ref. [9]. While the i-TS model applies up to 10 MPs, Fig. 1 is the strongest evidence confirming that track radii depend only on a single MP. This contradiction was revealed already in Ref. [1], nevertheless it has not been resolved up until now though the i-TS model has been widely used during the elapsed time.

In previous sections, a relationship was analyzed between apparently independent ion-induced track data obtained in experiments on various insulators. When $<s_e>$=constant holds, $R_e$ values are given by a universal function $\Theta(r)$, and they do not depend on MPs apart from $T_m$. It is revealed that this is the consequence of the formation of an identical temperature distribution $\Theta(r)=\Delta T(r,0)$ [1]. These results are in contradiction with previous ideas to a great extent, and certainly deserve further study. It is reasonable to revise previous estimates made by using those theories in cases when the deviation from $\Theta(r)$ are significant. In what follows, some considerations about the possible mechanism are discussed.

### III. SOME FEATURES OF THE UNDERLYING MECHANISM

Recently, various models used for the analysis of ion-induced track formation were reviewed [3]. All those models apart from ATSM are not coherent with i) - vii) in section II. Without going into details, it is evident, that neither of those models can lead to the formation of an ion-induced temperature distribution that does not depend on MPs. There must exist a further, presently unknown mechanism, which is responsible for the common features of track formation discussed earlier.

This must be a composition independent mechanism since the induced effect is identical at the given value of $<s_e>$ whatever is the composition of the irradiated insulator. However, this mechanism is efficient only in a narrow volume around the trajectory of the projectile where the amorphous tracks are formed (this has a radius of the order of 10 nm). When $S_e>S_{et}$ in an insulator, tracks are formed by each swift heavy ion. Thus this mechanism is activated at each impact of projectiles and evidently it does not control the cooling phase.

The parameter w=4.5 nm appears as the Gaussian width of the maximum temperature distribution $\Theta(r)$. Its relationship with basic length units [1] is an indication that it may be a key parameter of the underlying mechanism.

At high ion velocities, this new mechanism is responsible for the transfer to thermal energy of a relatively small fraction - about 17% - of the projectile energy (f=0.17). This value is derived unambiguously from the analysis of the threshold values of track formation $S_{et}$ in numerous insulators for E>8 MeV/nucleon and f=0.17 appears as a parameter in the expression of the Θ function, as well. When reducing the ion velocity in the range 2 MeV/nucleon< E <8 MeV/nucleon the VE is gradually superimposed on the composition independent mechanism and larger tracks are formed which are characterized with f>0.17. However, VE does not destroy the composition independent feature of the effect, as the LO and HI curves coincide in Fig. 1.

When tracks are studied experimental information is obtained only on that part of the irradiated solid where the actual processes can affect the track parameters. Usually, this is a cylinder with a radius of the order of 10 nm. This mechanism, which is activated by the irradiation of insulators by swift heavy ions is independent on the composition of the target within this cylindrical volume. With increasing distance from the trajectory, the validity of this statement gradually ceases and various other contributions may appear.

Unlike in insulators, the ion irradiation does not induce an identical temperature distribution in semiconductors. This is because the w parameter is an MP in semiconductors due to the higher mobility of charge carriers varying in a broad range. This makes large difference in the track formation in insulators and semiconductor. As a result, a broader thermal spike is formed in semiconductors in agreement with the prediction in Ref. 17. As w≠constant, the uniformity of track formation disappears here.

In semiconductors, VE is missing and f≈0.17 at any ion energy [18]. If the agreement of the efficiencies is not only an occasional coincidence, this may be an indication that the mechanism responsible for the uniformity in insulators might be active in semiconductors, as well. Thus in spite of the considerable differences, the same mechanism of track formation may be active in both types of solids. Further studies are necessary in this field.

Finally, we recall that in many experiments the initial track radii $R_{in}$ cannot be measured because of the fast processes of recrystallization or phase transformation. A possible solution of the problem is the estimation of the total track radius, similarly to the case of $Gd_2Ti_2O_7$. In Ref. 19 track diameters obtained from high-resolution TEM images were used for an estimate, which included contributions from the amorphous core and the disordered defect-fluorite shell. The origin of both phases is the molten region of the solid, and both phases modify the TEM contrast. These estimates for the total track radius in $Gd_2Ti_2O_7$ fit well to the $R_{in}$ data in Fig. 1.

In agreement with the prediction of ATSM, the track evolution proceeds along an $R_e^2$-$\ln(<s_e>/3\pi k w^2(T_m-T_i)$ curve (see Eq. (3)) without the application of any individual fitting parameter and MPs apart from $T_m$. This behavior is characteristic for cases when $R_{in}$ can be measured and according to recent results, it is also valid when only the total track radii are accessible, as in $Gd_2Ti_2O_7$ [6]. This supports the assumption that the application of the total track radii may provide a good approximation for $R_{in}$ in other solids, as well. Presently, a systematic study of total track radii has been reported only in $Gd_2Ti_2O_7$. Further experiments similar to those in Ref.[6] would be useful.

However, it would not be correct to claim that Eq. (3) is valid only in those insulators where $R_{in}$ values or total track radii are known, and it is not considered to be valid in all those insulators where $R_{in}$ cannot be measured or estimated. Namely, the functioning of the mechanism leading to composition independent, uniform temperature distribution (Eq. (3)) must not depend on whether $R_{in}$ can be measured or estimated in the given solid by applying the existing experimental techniques, or cannot. The validity of this mechanism extends to a much broader set of insulators.

The microscopic models proposed up to now cannot explain the common features of track formation leading to the relationship demonstrated in Fig. 1. Finding the right mechanism must solve the origin of the induced Gaussian temperature distribution $\Theta(R_e)$ with the width w=4.5 nm, and the simple relationship of the w parameter with basic length unit [1]. Besides the deeper understanding of the physical processes its application will lead to more reliable estimates of the properties of the ion-induced structures.

## IV. Conclusions

Experiments on insulators show that track radii $R_e$ depend only on a single MP - $T_m$ that proves unambiguously the validity of the thermal spike mechanism of track formation. A quantitative relationship exists between $R_e$ values induced in different insulators by different ions of different energies when $<s_e>$=constant. Up until now, this has been ignored in all theories of irradiation effects apart from ATSM. The consequences of the relationship between track radii are discussed in details. Several widely used assumptions are in contradiction with this relationship namely the origin of the VE, the effect of superheating, heat of fusion and band gap energy $E_g$ on the track size. The physical origin of this behavior is that a common composition independent temperature distribution $\Delta T(r,0)=\Theta(r)=(f<s_e>/3\pi kw^2)\exp\{r^2/w^2\}$ is induced in insulators close to the trajectory of the projectile that can be determined experimentally by track measurements. The Gaussian width of the distribution w=4.5 nm is in a simple relationship with basic length units: the Bohr radius, the Compton wavelength and the classical electron radius. The underlying mechanism does not depend on the composition of the solid and only the $0.17S_e$ fraction of the deposited electronic energy is controlled by it. But this invariance may gradually disappear with increasing distance from the track. The differences between the behavior of semiconductors and insulators are also discussed. Further studies are needed to reveal the presently unknown mechanism of energy deposition related to track formation.

## Data Availability

The datasets analyzed during the current study are available from the corresponding author on reasonable request.